# Like charge attraction in confinement: Myth or truth?

Jörg Baumgartl[1], Jose Luis Arauz-Lara[2], and Clemens Bechinger[1]

[1] 2. Physikalisches Institut, Pfaffenwaldring 57, 70550 Stuttgart, Germany
[2] Instituto de Física, Universidad Autónoma de San Luis Potosí, Alvaro Obregón 64, 78000 San Luis Potosí, Mexico

It is general wisdom that likely charged colloidal particles repel each other when suspended in liquids. This is in perfect agreement with mean field theories being developed more than 60 years ago. Accordingly, it was a big surprise when several groups independently reported long-ranged attractive components in the pair potential $U(r)$ of equally charged colloids. This so-called *like-charge attraction* (LCA) was only observed in thin sample cells while the pair-interaction in unconfined suspensions has been experimentally confirmed to be entirely repulsive. Despite considerable experimental and theoretical efforts, LCA remains one of the most challenging mysteries in colloidal science. We experimentally reinvestigate the pair-potential $U(r)$ of charged colloidal particles with digital video microscopy and demonstrate that optical distortions in the particle's images lead to slightly erroneous particle positions. If not properly taken into account, this artefact pretends a minimum in $U(r)$ which was in the past misleadingly interpreted as LCA. After correcting optical distortions we obtain entirely repulsive pair interactions which show good agreement with linearized mean field theories.

A controversial debate in colloidal science has been launched in 1994 when Kepler and Fraden reported an unusual long-ranged attractive component in the pair potential of charged colloidal particles [1]. This so-called *like-charge attraction* (LCA) was only observed in thin cells (typical sample height H < 10μm) while the pair-interaction in bulk suspensions has been experimentally confirmed to be entirely repulsive in agreement with Poisson-Boltzmann theories [2-4]. Therefore, it was speculated that the confining plates are responsible for the deviations from theory. Soon after its initial observation LCA was also observed by other authors [5-11] which then provoked considerable theoretical interest in this phenomenon. In the meantime it has been rigorously proven that the observed attraction can not be explained within the framework of mean field theories, irrespective of whether the particles are suspended in bulk or in confinement [12-14]. Several other approaches beyond Poisson-Boltzmann have been proposed as the origin of confinement-induced attraction. While correlation effects of the electrolyte indeed can lead to a short-ranged attraction of negatively charged colloids, none of the existing theories can account for the long-ranged attraction as observed in experiments [15,16]. Accordingly, even after more than ten years of considerable research, LCA is still one of the most challenging unsolved mysteries in colloidal science.

In this study we reinvestigate the pair potential $U(r)$ of charged colloidal particles in thin sample cells. Contrary to previous experiments where $U(r)$ was derived from the pair correlation function $g(r)$ in semi-dilute suspensions, here we determine the pair potential directly by measuring the probability distribution of two silica spheres. This is of particular advantage because it avoids the need of inversion procedures (Ornstein-Zernike, inverted Monte-Carlo techniques) which not always produce reliable results for $U(r)$ [17,18]. Like in previous studies on LCA, we also determined particle positions with digital video microscopy which is capable to localize particle centres with accuracy down to below 20nm. In a recent paper, we demonstrated that overlapping particle images lead to small but systematic optical distortions which then result in erroneously determined particle positions [19]. If not properly taken into account, this optical artefact pretends a minimum in U(r) which exactly resembles the features previously interpreted as LCA. The purpose of this paper is, to add more experimental evidence which demonstrates that LCA is the result of uncorrected data. First, we briefly discuss the origin of optical distortions in digital video microscopy and show the influence of uncorrected data on measured pair potentials. To demonstrate that a minimum in U(r) is not related to particle confinement we also measured pair potentials in thick sample cells and demonstrate that even without confinement artificial minima can occur under certain conditions. Finally, we suggest a method how optical artefacts can be avoided in future experiments.

Interaction potentials between colloids have been obtained by subjecting two particles to a well-defined radially symmetric light potential which was created by a slightly defocused circular polarized Gaussian laser beam ($TEM_{00}$, $\lambda$ = 532 nm). The light pressure from the vertically incident laser beam confines the particles irrespective of the cell height to two dimensions [20]. From the particle trajectories as measured by digital video microscopy (for a review see e.g. [21]) we obtain their relative distance distribution P(r) which then leads to the pair-potential $P(r)=P_0 exp(-\{U_{ext} +U(r)\}/k_B T)$ with $U_{ext}$ the potential induced by the laser tweezers. In the central region which is sampled by the particles we confirmed that $U_{ext}$ is approximated very well by a parabolic potential. This shape is particularly convenient because in parabolic potentials the centre of mass motion decouples from the relative particle motion and thus allows to obtain $U_{ext}$ and U(r) simultaneously (otherwise the external potential must be first determined by the probability distribution of a single colloidal particle inside the laser trap [4]). All pair potentials as shown in the following were performed at room temperature and reproduced for different laser intensities up to about 50mW resulting in different optical trapping strengths. This was done to rule out possible light-induced effects such as optical binding [22].

The experiments were performed with highly charged silica particles of diameter $\sigma$ = 1.5 ± 0.08 µm which were confined in thin sample cells. The fabrication of thin cells has been described in detail elsewhere [8]. Briefly, bidisperse suspensions containing the above silica particles and a small amount of slightly larger polystyrene particles ($\sigma$ = 1.96 µm) were dialyzed against pure water for several weeks to obtain good deionization. A small amount of this mixture was confined between two clean glass plates which were uniformly pressed on top of each other until the distance H between the glass plates is determined by the larger particles. Accordingly, the larger particles were randomly distributed across the sample while the smaller ones diffuse through the cell. Afterwards the system was sealed with epoxy resin which yielded stable conditions over several weeks. The silica spheres form a monolayer with their centres about 800-900 nm above the lower glass wall. Out-of-plane excursion can be estimated to be less than 100nm [23]. In previous experiments it has been demonstrated that the pair potential of these particles exhibits a long-range attraction of several tenths of $k_B T$ at r ≈ 1.5 $\sigma$ [5,9]. The colloids were illuminated from above under Köhler conditions with incoherent white light and imaged with an oil immersion objective (100x, NA=1.25) from below onto a CCD camera which was connected to a computer. Particle positions were determined online with an imaging processing software (upper inset of Fig.1).

The open symbols in Fig.1 shows the measured pair potential in a thin sample cell. In addition to the strong repulsion close to contact, we find a shallow minimum with a depth of approximately 0.2 $k_B T$ at r ≈ 1.3$\sigma$ which is in excellent agreement with the characteristic features of LCA [1,5,6,11]. In the following we demonstrate that the occurrence of such a minimum must not be interpreted as an attractive component in the pair potential but is the result of overlapping colloidal images which lead to slightly wrong particle distances obtained by digital video microscopy.

We want to start our discussion by recalling that images of colloidal particles are typically much larger than their geometrical size [19]. This is well-known for transparent spherical particles and also confirmed by numerical calculations [24]. To illustrate this effect, we show as insets in Fig.2a-d snapshots of two particles for distances between 1.25σ and 2.58σ. The focusing conditions which influence the optical appearance of the colloid has been adjusted to obtain a particle image as a bright central spot surrounded by a dark ring. These settings are of particular advantage for identifying particle positions in video microscopy and have been also applied by other groups reporting LCA [9,25]. From the corresponding intensity cross-sections in Fig.2a-d we determine the spatial extension of a particle image (defined by the outer diameter of the darker ring) to approx. 2σ. Obviously the particle image is blurred over almost twice the particle diameter. As a consequence, particle images start to overlap when particle centres are closer than 2σ. This overlap unavoidably causes a small but systematic asymmetry of the intensity distribution around the particle centres which is most clearly seen in Fig.2d. Because particle positions are obtained by calculating the intensity weighted centroids above a threshold, any change in the intensity distribution must necessarily lead to a deviation between the true and the measured particle distance. Although this has been already realized by other authors [26], at that time neither a quantitative analysis nor the consequences for experimentally measured pair potentials were investigated.

To study the influence of overlapping particle images on their measured distance, we performed precise position measurements of two colloidal spheres. As an immobile reference position we selected an isolated particle which was irreversibly stuck (due to van-der Waals forces) to the substrate and determined its probability distribution. Due to the finite experimental resolution, the theoretically expected delta-like probability distribution is broadened by about 20 nm and the maximum is identified as the *true particle position* $(\bar{x}_{ref}, \bar{y}_{ref})$. Next, we laterally approached a non-sticking particle with a laser trap and determined for each video frame the *momentarily* position of the immobile reference particle $(x_{ref}(t), y_{ref}(t))$ and that of the mobile one $(x(t), y(t))$. Note, that in contrast to $(\bar{x}_{ref}, \bar{y}_{ref})$ the position $(x_{ref}(t), y_{ref}(t))$ is obtained in the presence of another particle. From these positions we calculated the momentary particle distance $r(t) = \sqrt{(x_{ref}(t) - x(t))^2 - (y_{ref}(t) - y(t))^2}$. As a quantitative measure of how overlapping particle images modify their measured distance we introduce

$\Delta r(t) = r(t) - \sqrt{[x(t) - \bar{x}_{ref}]^2 + [y(t) - \bar{y}_{ref}]^2}$. It is immediately obvious that this expression is essentially zero if the position of the isolated reference particle is not modified by the presence of another close particle, i.e. $(\bar{x}_{ref}, \bar{y}_{ref}) = \langle (x_{ref}(t), y_{ref}(t)) \rangle$ where the outer bracket corresponds to time averaging. Fig.3a shows the result of $\Delta r(t)$ when the mean particle distance is large enough ($\approx 10\sigma$) to avoid any overlap of the particle's images. As expected, $\Delta r(t)$ scatters within our experimental resolution around zero and thus demonstrates at that large distances video microscopy yields accurate positional information. However, when we reduce the particle distance r<2σ where particle images overlap, $\Delta r(t)$ looks rather different Fig.3c). In addition to almost doubled fluctuations, the corresponding histogram is asymmetric with its centroid shifted to the right (Fig3d). Obviously, at those distances, the position measurement is affected by the presence of another particle. When replotting the randomly fluctuating $\Delta r(t)$ in Fig.3c as $\Delta r(r)$ we observe a characteristic distance dependence which is plotted in Fig.4

For symmetry reasons the image distortion affects both particles identically, therefore the *true particle distance* $r_t$ is simply given by $r_t = r - 2\Delta r$. For clarity, in the following we will refer to r as the *measured distance*. From the $\Delta r(r)$ plot we immediately see that for r < 1.15 σ the particles are closer than they appear while for 1.15 σ < r < 1.8 σ they are further apart than suggested by the intensity weighted centroids. Having obtained a relationship between r and $r_t$, we finally obtain the corrected distribution $P_t(r_t) = P(r)\frac{dr}{dr_t}$ which then leads to the corrected pair potential. Since $\frac{dr}{dr_t}$ strongly depends on the particle distance, the corrected potential is not simply shifted but undergoes a more complicated transformation after which the minimum disappeared (closed symbols in Fig.1). The resulting potential can be successfully described with a screened Coulomb potential [27,28]

$$U(r) = (Z^*)^2 \lambda_B \left( \frac{\exp(\kappa\sigma/2)}{1+\kappa\sigma/2} \right)^2 \frac{\exp(-\kappa r)}{r}$$

where $Z^* \approx 8000 \pm 4000$ is the effective colloidal charge, $\kappa^{-1} \approx 10 \pm 5$ nm the Debye screening length, and $\lambda_B = 0.72$nm the Bjerrum length in water. Such entirely repulsive pair interactions are in perfect agreement with theoretical predictions and thus confirm the validity of one of the most basic interactions in charge-stabilized colloidal suspensions even under confined conditions.

So far, we demonstrated that overlapping colloidal images lead to erroneous particle distances which – if not taken into account – can lead to long-ranged attractive components in U(r). If LCA is really the result of falsely interpreted video images, why the above discussed optical artefacts are only relevant in thin sample cells while pair potentials in unconfined systems were concordantly reported to agree with a screened Coulomb potential [3,23,25] ? This is easily understood by recalling that the Debye screening length in charged colloidal suspensions is given by $\kappa^{-1} \propto (z_s c_s + z_c c_c)^{-1/2}$ where $z_{s,c}$ and $c_{s,c}$ are the valency and the concentrations of the salt ions and that of the counter ions, the latter released from any surfaces in the system. While in thick sample cells the main source of counter ions are colloidal surfaces, with decreasing cell thickness H counterions released from the surfaces of the sample cell become increasingly important. As a result, with decreasing H the screening length and therefore the minimal particle distance which is sampled by the colloids in thermal equilibrium decrease [25]. Because particle image overlap occurs only for r<2σ this makes interaction measurements in confined conditions particularly susceptible for the above artefacts. In contrast, particle distances in U(r) measurements of unconfined samples hardly fall below 2σ [3,23,25] where video microscopy yields correct potentials. This interpretation is also consistent with an observation made in the first paper on LCA where it was reported that even under confinement, a minimum is only observed for sufficiently high ionic strength [1].

To demonstrate that a minimum in U(r) is not at all related to particle confinement we measured the pair potential in a thick cell under different salt concentrations. Fig.5a shows U(r) of the same silica particles as used above for H=200μm under rather deionized conditions (ionic conductivity ≈ 0.8 μS/cm). Particle distances below 2σ are hardly sampled and the data agree well with a screened Coulomb potential (solid line). From a fit we obtain $Z^* \approx 15900 \pm 1200$ and $\kappa^{-1} \approx 190 \pm 5$nm. Increasing the salt concentration (corresponding to an ionic conductivity of 4 μS/cm) leads to a shift of the potential towards smaller distances. In addition, however, the uncorrected potential exhibits a pronounced minimum (Fig.5b). Again, the depth and position of this minimum is in excellent agreement with the characteristics of LCA but is here interestingly observed *in the absence of confinement*. Applying the correction procedure as described above, eventually leads to perfect agreement with a screened Coulomb potential with $Z^* \approx 18700 \pm 1500$ and $\kappa^{-1} \approx 55 \pm 5$nm (Fig.5c). Finally we calculated the bare charge $Z_{bare}$ obtained from $Z^*$ and $\kappa^{-1}$ using an analytical expression based on the Poissson-Boltzmann mean-field theory [29] and obtain $Z_{bare}$ = 19210 and 19275 for the low and high ionic concentrations, respectively. The almost perfect agreement is an independent consistency check for the corrected U(r) and demonstrates that our correction leads to reasonable parameters.

Having identified the pitfalls in video microscopy, finally we want to address the question how one can avoid optical artefacts rather than correcting them. From the above it is clear that one has to avoid overlapping particle images which are the real source of the problem. One approach is to use core-shell particles where e.g. the core is labelled with a fluorescent dye or where the shell is matched exactly to the solvent [30]. For sufficiently large shell thicknesses, in both cases the particle image can be smaller than its geometrical size. This would avoid artefacts even when the particles are in physical contact. The insets of Fig.6 shows a highly diluted system of silica spheres with a fluorescent (fluorescein isothiocyanate, FITC)) core. The core diameter was 400nm and the total radius 1.4µm. In the fluorescence image (upper inset) the particle appears even smaller than their geometrical size. As expected, in this case the $\Delta r$ vs. r curve (Fig.6a) scatters symmetrically around zero and shows no particular structure (note that the increased noise is due to the smaller particle image [21]). This suggests the use of such core-shell particles in future studies to avoid optical distortions. In comparison, white light illumination leads again to images which are blurred over almost twice the particle diameter and thus leads to optical distortions and the characteristic shape of $\Delta r(r)$ (Fig.6b, lower inset).

In summary, our experiments demonstrate that overlapping particle images lead to erroneous particle distances as obtained by digital video microscopy. If not taken into account, these artefacts lead to an apparent attractive component in the pair interaction of negatively charged colloids. Because this effect has not been considered in previous studies who claimed like-charge attraction of confined colloidal suspensions, our results provide a natural explanation for this highly controversially discussed phenomenon. In order to avoid optical artefacts in future studies with small particles distances, we suggest the use of colloids with fluorescent cores to avoid overlap of particle images.

**Acknowledgements**

It is a pleasure to thank C.v. Kats and A.v. Blaaderen for generous supply with fluorescent core-shell particles. We acknowledge technical support from A. Ramirez-Saito during the preparation of thin sample cells and helpful comments from L. Helden, V. Blickle and D. Babic.

**Figure Captions**

**Fig.1** Pair potential of two 1.5µm silica particles in a thin sample cell (H= 1.96µm). The open symbols correspond to the uncorrected data while the closed data points are obtained after optical artefacts were corrected. Upper inset: schematic view of the experimental setup with a laser tweezers (grey) and white light illumination (arrows). The laser is blocked with an optical filter between the sample and the microscope objective. Lower inset: Video snapshot taken during the measurement.

**Fig.2** Intensity cross sections of two particle images for different measured distances $r$ : (a) $r = 2.58\sigma$ (b) $r = 1.93\sigma$ (c) $r = 1.67\sigma$ and (d) $r = 1.25\sigma$. Insets: Corresponding video snapshots.

**Fig.3** $\Delta r(t)$ between a fixed and a free fluctuating particle in a thin cell (H = 1.96µm) as determined by video microscopy (a) for large and (d) small particle distances. (b,d) corresponding histograms of the measured distance fluctuations.

**Fig.4** Experimentally determined $\Delta r(r)$ as obtained from the data plotted in Fig.3c. The solid line corresponds to the smoothed data.

**Fig.5** Uncorrected pair potentials for H = 200μm and (a) ionic conductivity of $4\,\mu S/cm$ and (b) $0.8\,\mu S/cm$. After correcting the data in (b) for imaging artefacts we obtain (c). The solid lines are fits to a Yukawa potential, the dotted line is a guide to the eye. Inset: $\Delta r$ vs. $r$ (grey). The solid line is a polynomial fit.

**Fig.6** $\Delta r$ vs. $r$ for core-shell particles with a fluorescent core: (a) fluorescent illumination (the data are shifted in vertical direction by 0.075 and (b) bright-field illumination. The insets display typical snapshots.

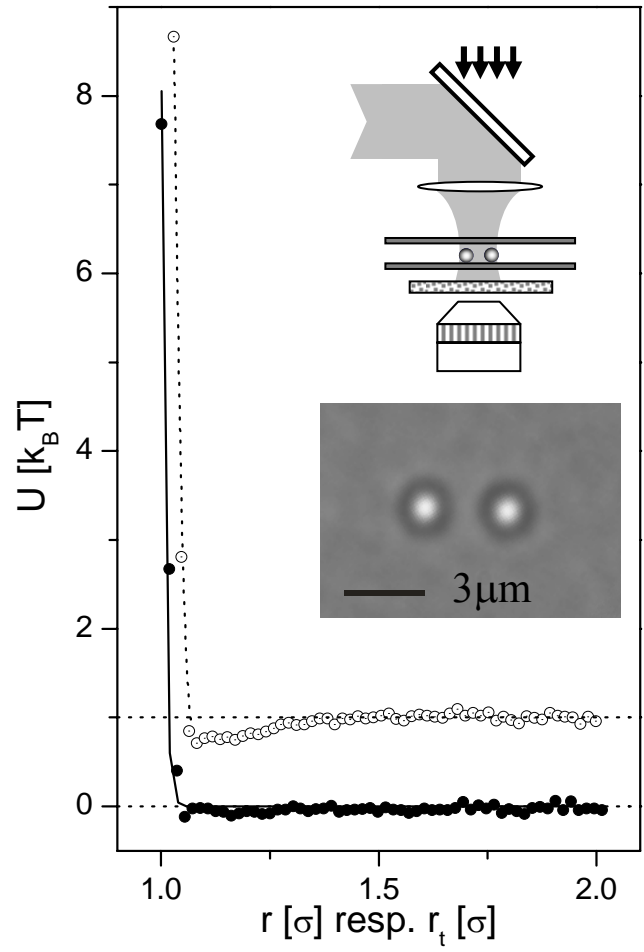

Fig.1

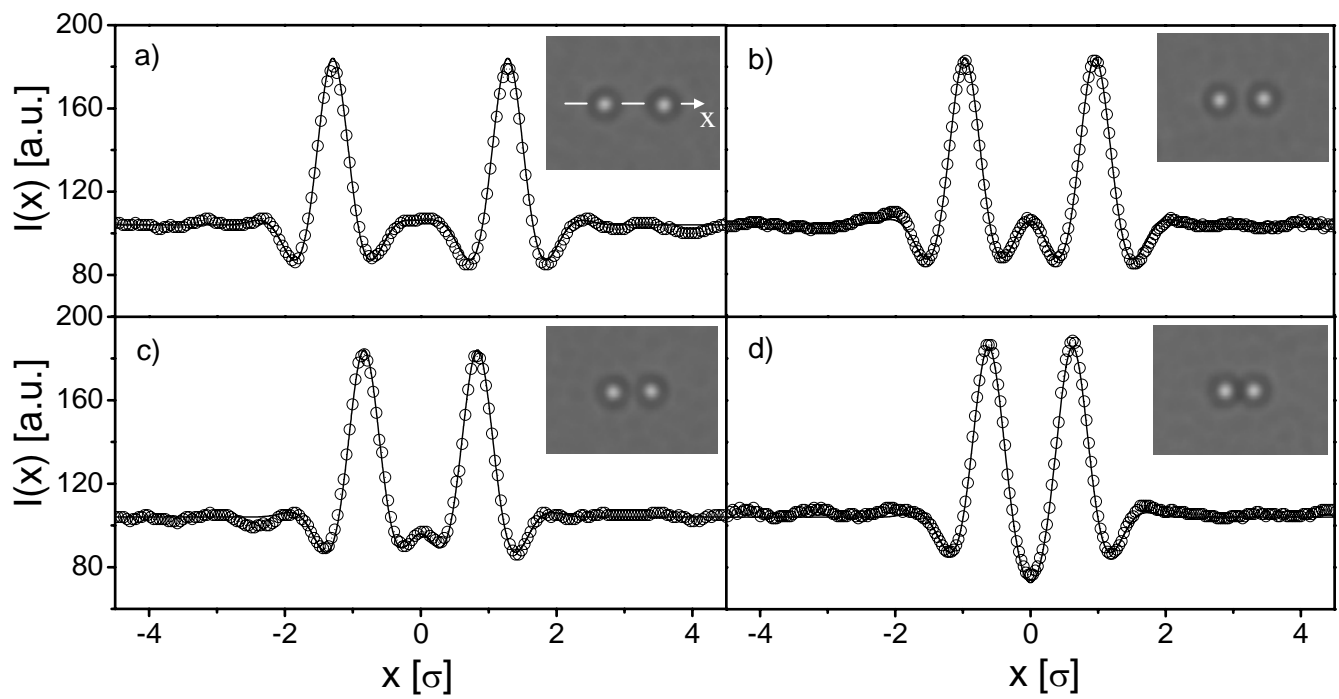

Fig.2

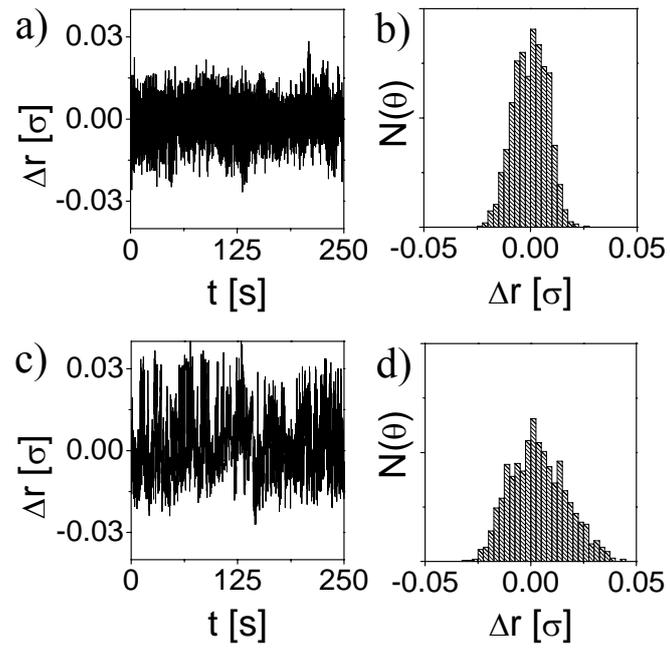

Fig.3

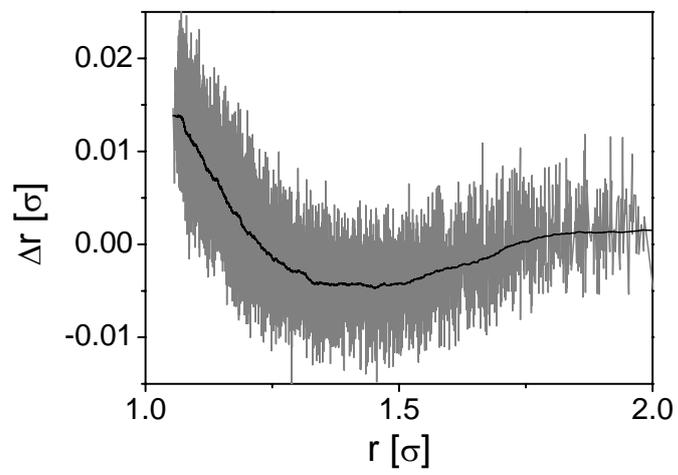

Fig.4

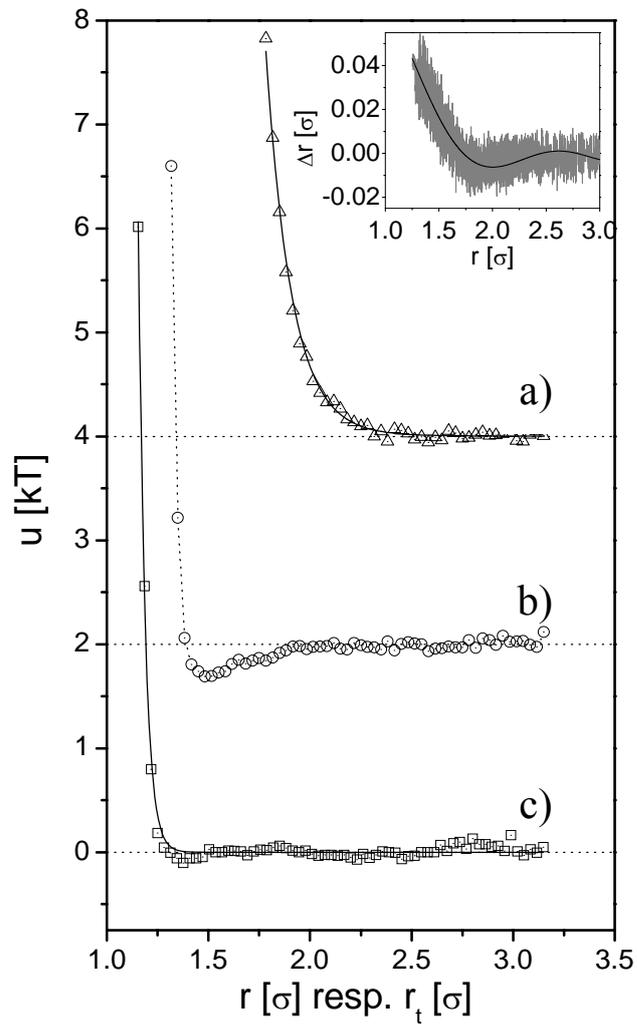

Fig.5

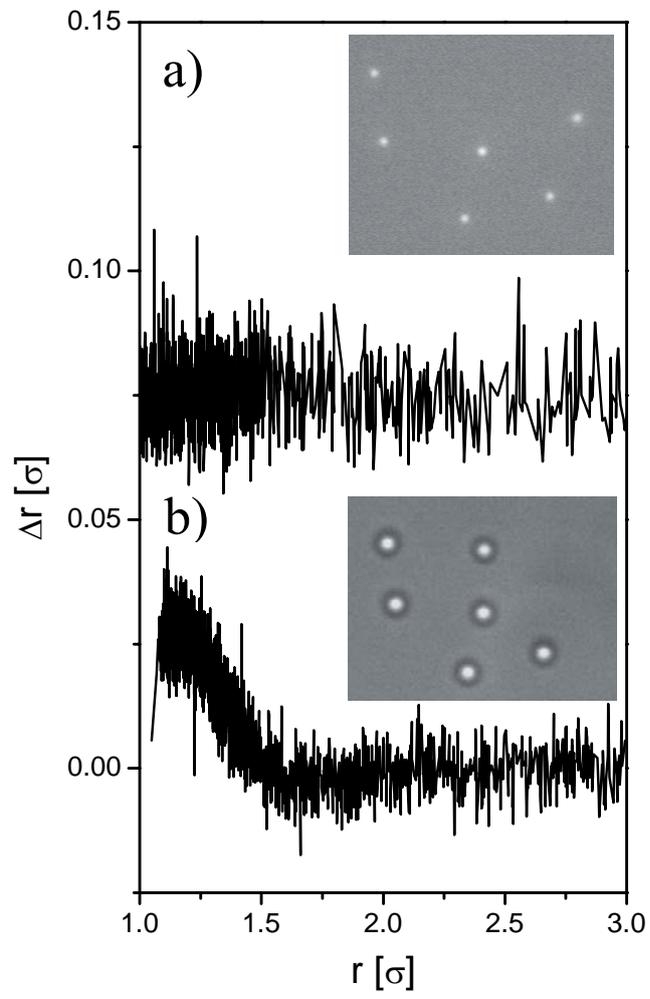

Fig.6